\documentclass[aps,pre,reprint,nofootinbib]{revtex4-1}
\usepackage{amsmath,amssymb,amsthm}
\usepackage{latexsym, graphicx}
\usepackage[pdftex,colorlinks]{hyperref}

\newcommand{\bra}[1]{\left\langle #1 \right|}
\newcommand{\ket}[1]{\left| #1 \right\rangle}

\theoremstyle{definition}

\theoremstyle{definition}

\begin{document}

\title{Initial state preparation with dynamically generated system-environment
correlations}
\author{C. H. Fleming, B. L. Hu}
\affiliation{Joint Quantum Institute and Department of Physics, University of Maryland, College Park, Maryland 20742}
\author{Albert Roura}
\affiliation{Max-Planck-Institut f\"ur Gravitationsphysik (Albert-Einstein-Institut),
Am M\"uhlenberg 1, 14476 Golm, Germany}

\begin{abstract}
The dependence of the dynamics of open quantum systems upon initial correlations between the system and environment is an utterly important yet poorly understood subject. For technical convenience most prior studies assume factorizable initial states where the system and its environments are uncorrelated, but these conditions are not very realistic and give rise to peculiar behaviors. One distinct feature is the rapid build up or a sudden jolt of physical quantities immediately after the system is brought in contact with its environments. The ultimate cause of this is an initial imbalance between system-environment correlations and coupling. In this note we demonstrate explicitly how to avoid these unphysical behaviors by proper adjustments of correlations and/or the coupling, for setups of both theoretical and  experimental interest. We provide simple analytical results in terms of quantities that appear in linear (as opposed to affine) master equations derived for factorized initial states.
\end{abstract}

\maketitle


\section{Open system initial correlations}
An open quantum system is a quantum system `S' that interacts with some environment `E' whose degrees of freedom have been coarse grained (colloquially, `traced out' or `integrated over' ). The unitary evolution of the combined  system + environment, `C' (for combined or closed), is generated by the Hamiltonian
\begin{equation}
\mathbf{H}_\mathrm{C} \equiv \mathbf{H}_\mathrm{S} + \mathbf{H}_\mathrm{E} + \mathbf{H}_\mathrm{I} \, ,
\end{equation}
where $\mathbf{H}_\mathrm{I}$ denotes the system-environment interaction.
Specifying the initial of the combined system is necessary to determine the open-system dynamics.
The most common choice is to assume factorized initial states for the system and environment%
\footnote{Even after resolving issues of renormalization, the system will typically be displaced by a finite amount of the order of the induced damping within a very short time of the order of the inverse UV cutoff of the environment.}
\begin{eqnarray}
\boldsymbol{\rho}_\mathrm{C}(0) &=& \boldsymbol{\rho}_\mathrm{S}(0) \otimes \boldsymbol{\rho}_\mathrm{E}(0) \, , \\
\boldsymbol{\rho}_\mathrm{E}(0) &=& \frac{1}{Z_\mathrm{E}(\beta)} e^{-\beta \, \mathbf{H}_\mathrm{E}} \, .
\end{eqnarray}
where $Z_\mathrm{E}(\beta)$ denotes the partition function of the free (noninteracting) environment and $T=1/\beta$ is the temperature of the environment, which acts here as a thermal reservoir.

When considering environments with a large number of high-frequency modes and characterized by a UV frequency cutoff $\Lambda$,
such a factorized initial state (chosen for mathematical simplicity) unfortunately engenders unphysical behavior such as a sudden jolt in physical quantities near the initial time (this was analyzed in some detail in Ref.~\cite{HPZ92}) or spurious cutoff sensitivity of certain system correlators (see Ref.~\cite{HRV04} and references therein).
This kind of initial conditions assumes that an uncorrelated system and environment are instantaneously coupled with non-vanishing strength.  The pathological behavior arises because
the factorized initial state contains a number of highly excited energy states of the full  Hamiltonian (including the interacting Hamiltonian),
even when the initial reduced states of the system and environment are not highly excited in the free theory, and it is a reflection of the high-frequency modes of the environment quickly becoming correlated with the system within a time of order $1/\Lambda$.

The next most common choice of initial state (see Ref.~\cite{Grabert88}) has been to consider system deformations or measurements of the global equilibrium state of the combined system `C', with density matrix
\begin{equation}
\boldsymbol{\rho}_\mathrm{C}(0) = \sum_n \mathbf{O}'_n \, \frac{1}{Z_\mathrm{C}(\beta)} e^{-\beta \, \mathbf{H}_\mathrm{C}} \, \mathbf{O}_n \, ,  \label{eq:PrepP}
\end{equation}
where the $\mathbf{O}$ and $\mathbf{O'}$ operators are restricted to act on the system.
However, this method still gives rise to jolts for sufficiently general deformations or measurements \cite{Romero97}, which can be understood as a consequence of altering the state of the system instantaneously \cite{Anglin97}.

To cure or tame these drastic effects, especially in the context of linear systems, the following procedure has been suggested: a) force the system by a constant amount, b) wait for it to relax into the displaced equilibrium state, and then c) release the force \cite{Grabert88}. Alternatively and in order to generate interesting coherent superposition states for the system, one can start with the equilibrium state of the combined system and act on the system, but for a non-vanishing time \cite{Anglin97}.
Essentially we view the problem as an imbalance between initial correlations and initial coupling strength; the imbalance can be countered on either side.
We also believe that the most natural resolution should be a dynamical preparation which relies upon equilibration \cite{Grabert88,Breuer01b} followed by an additional preparation of the system for a finite time \cite{Anglin97}. Our key contribution is showing that this can be achieved while still taking advantage of the simpler analytical results obtained when deriving the master equation for a factorized initial state, without the need to introduce inhomogeneous terms and an affine master equation \cite{Breuer01b}.

In the next section we will briefly discuss the perturbative open-system master equation which we will use to approach these issues.
Then in Secs.~\ref{sec:couple} and \ref{sec:Hamiltonian} we will provide resolutions based, respectively, on balancing the coupling and the correlations.

\section{Open-system dynamics}
In the time-local representation (also called the convolutionless or Markovian representation), the dynamics of the reduced density matrix of the system $\boldsymbol{\rho}$ can be expressed with a quantum Liouville equation
\begin{eqnarray}
\frac{d}{dt} \boldsymbol{\rho}(t) &=& \boldsymbol{\mathcal{L}}(t) \, \boldsymbol{\rho}(t) \, \label{eq:TCME},
\end{eqnarray}
for any factorized initial condition.
As a perturbative approximation, $\boldsymbol{\mathcal{L}}(t)$ is expanded in powers of the system-environment interaction $\mathbf{H}_\mathrm{I}(t)$ and truncated to some order.
(We momentarily consider full time dependence in this section as the more general relations will be necessary for our techniques.)
Such perturbative master equations can be derived in a variety of  ways \cite{Kampen97,Breuer01,Strunz04}
and find application in many branches of physics and chemistry \cite{Pollard97,Carmichael99,Breuer02,Kampen07}.
The expansion of $\boldsymbol{\mathcal{L}}(t)$ will then take the form
\begin{eqnarray}
\boldsymbol{\mathcal{L}}(t) &=& \sum_{k=0}^\infty  \boldsymbol{\mathcal{L}}_{2k}(t) \, ,\label{eq:PerturbExp} \\
\boldsymbol{\mathcal{L}}_{0} \, \boldsymbol{\rho} &=& \left[ -\imath \, \mathbf{H}_\mathrm{S}(t) , \boldsymbol{\rho} \right] \, ,
\end{eqnarray}
where $ \boldsymbol{\mathcal{L}}_{2k} = \mathcal{O}(\mathbf{H}_\mathrm{I}^{2k})$ and to zeroth-order the system is driven in a unitary manner by its Hamiltonian $\mathbf{H}_\mathrm{S}(t)$.
We take the expansion to be even as we only consider Gaussian noise, which is symmetric.
A Gaussian noise distributional is necessary for higher-order perturbation theory to be non-secular in the late-time limit \cite{QOS},
though at second order one is effectively truncating the noise cumulants in a manner consistent with Gaussian noise.

The linear master equation is derived under the general assumptions of a factorized initial state and an expansion of the interaction as a sum of separable operators:
\begin{equation}
\mathbf{H}_\mathrm{I}(t) = \sum_n \mathbf{L}_n(t) \otimes \mathbf{l}_n(t) \, , \label{eq:Hint}
\end{equation}
where $\mathbf{L}_n(t)$ and $\mathbf{l}_n(t)$ are system and environment operators respectively.
The environment coupling operators $\mathbf{l}_n(t)$ will typically be collective observables of the environment, with dependence upon very many modes.

Using the notation of Ref.~\cite{QOS}, the second-order master equation can be expressed as
\begin{equation}
\boldsymbol{\mathcal{L}}_2 \{ \boldsymbol{\rho} \} \equiv \sum_{nm} \left[ \mathbf{L}_n, \boldsymbol{\rho} \, (\mathbf{A}_{nm}\! \diamond \mathbf{L}_m)^\dagger - (\mathbf{A}_{nm}\! \diamond \mathbf{L}_m) \, \boldsymbol{\rho} \right] \, , \label{eq:WCGME}
\end{equation}
where the $\mathbf{A}$ operators and $\diamond$ product define the second-order operator
\begin{equation}
(\mathbf{A}_{nm}\! \diamond \mathbf{L}_m)(t) \equiv \int_0^t \!\! d\tau \, \alpha_{nm}(t,\tau) \, \left\{ \mathbf{G}_0(t,\tau) \, \mathbf{L}_m(\tau) \right\} \, . \label{eq:WCOG}
\end{equation}
Here  $\mathbf{G}_0(t,\tau): \boldsymbol{\rho}(\tau) \to \boldsymbol{\rho}(t)$
is the free-system propagator, which for a constant Hamiltonian $\mathbf{H}$ is given by
\begin{equation}
\mathbf{G}_0(t,\tau) \, \boldsymbol{\rho}= e^{-\imath(t-\tau) \mathbf{H}} \, \boldsymbol{\rho} \, e^{+\imath(t-\tau) \mathbf{H}} \, ,
\end{equation}
while $\alpha_{nm}(t,\tau)$ are the environment correlation functions defined by
\begin{equation}
\alpha_{nm}(t,\tau) = \left\langle \underline{\mathbf{l}}_n\!(t) \, \underline{\mathbf{l}}_m\!(\tau) \right\rangle_\mathrm{E} \, , \label{eq:alpha}
\end{equation}
where $\underline{\mathbf{l}}_n\!(t)$ represents the time-evolving $\mathbf{l}_n$ in the interaction (Dirac) picture.
In general the correlation function is Hermitian and positive definite.
For constant coupling to any stationary environment, the correlation function will also be stationary, $\boldsymbol{\alpha}(t,\tau)=\boldsymbol{\alpha}(t\!-\!\tau)$.
Furthermore, for a thermal environment the correlation function will satisfy the KMS relation \cite{Kubo57,Martin59}:
\begin{equation}
\tilde{\boldsymbol{\alpha}}(\omega) = \tilde{\boldsymbol{\alpha}}^*(-\omega) \, e^{-\beta \omega} \, ,
\end{equation}
where $\tilde{\boldsymbol{\alpha}}(\omega) = \int_{-\infty}^{+\infty} dt \, \boldsymbol{\alpha}(t) \, e^{-\imath \omega t}$ denotes the Fourier transform.

From this perspective, the mathematical cause of the initial jolt becomes clear.
For constant Hamiltonians and an initially stationary environment, the second-order operator obeys the relation
\begin{equation}
\frac{d}{dt}(\mathbf{A}_{nm}\! \diamond \mathbf{L}_m)(t) = \alpha_{nm}(t) \, \left\{ \mathbf{G}_0(t) \, \mathbf{L}_m \right\} \, ,
\end{equation}
which can be extremely large near the initial time when considering an environment with a sufficient amount of high frequency modes (such as low-temperature ohmic and supra-ohmic environments) since $\boldsymbol{\alpha}(t)$ is typically a very localized distribution in those cases.
For a finite but large cutoff $\Lambda$, the correlation function becomes of order $\Lambda$ for a time of order $1/\Lambda$.

\section{Coupling switch-on}
\label{sec:couple}
One method for balancing the initial coupling between the system and environment with their initial lack of correlation, is to turn on the coupling slowly with a time-dependent interaction such as
\begin{equation}
\mathbf{H}_\mathrm{I} = \theta_\mathrm{s}(t) \sum_n \mathbf{L}_n \otimes \mathbf{l}_n \, , \label{eq:Hint}
\end{equation}
where $\theta_\mathrm{s}(t) : [0,\infty) \to [0,1)$ is a smooth switch-on function with a characteristic timescale $\tau_\mathrm{s}$, which vanishes at the initial time and becomes (effectively) one for times longer than $\tau_\mathrm{s}$.
To some extent, this was considered for linear systems in Ref.~\cite{QBM}.

Such a time-dependent interaction is equivalent to employing the second-order operator
\begin{equation}
(\mathbf{A}_{nm}\! \diamond \mathbf{L}_m)(t) = \theta_\mathrm{s}(t) \! \int_0^t \!\!\! d\tau \, \theta_\mathrm{s}(t\!-\!\tau) \, \alpha_{nm}(\tau) \left\{ \mathbf{G}_0(\tau) \, \mathbf{L}_m \right\} \, , \label{eq:WCOG}
\end{equation}
for otherwise constant couplings and Hamiltonians.
Therefore, any initial jolt due to the localized nature of $\boldsymbol{\alpha}(t)$ will be suppressed by $\theta_\mathrm{s}(t)$ as long as $\tau_\mathrm{s} \gg 1/\Lambda$.
\begin{figure}[h]
\centering
\includegraphics[width=7cm]{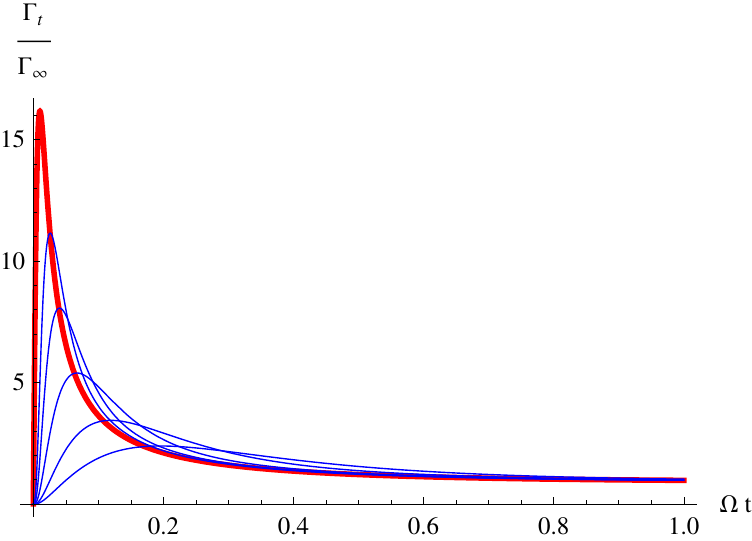}
\caption{Zero-temperature, ohmic decay rate for the \textcolor{red}{$\bullet$ instantaneously coupled} and \textcolor{blue}{$\cdot$ gradually coupled} initial states of a two-level system with exponential cutoff frequency $\Lambda = 100 \, \Omega$.
In this case the switch-on function is exponential, $\theta_\mathrm{s}(t) = 1-e^{-t/\tau_\mathrm{s}}$,
and the switch-on times $\tau_\mathrm{s}$ are chosen to take the values $1/\Lambda$, $2/\Lambda$, $4/\Lambda$, $8/\Lambda$, $16/\Lambda$. }
\label{fig:switched}
\end{figure}
As can be seen in Fig.~\ref{fig:switched}, the cutoff-frequency jolts are essentially replaced by jolts of frequency $\mathrm{min}[\Lambda,1/\tau_\mathrm{s}]$ and amplitude proportional to the same value.
This approach provides a useful way of generating initial system-environment correlations when $\tau_\mathrm{s}$ is much larger than $1/\Lambda$ but smaller than any other relevant timescales (such as the system frequencies). Furthermore, even if a mild jolt is still present, the important point is that it is cutoff insensitive (for fixed $\tau_\mathrm{s}$ and sufficiently large $\Lambda$).

\section{Dynamically prepared initial states}
\label{sec:Hamiltonian}
Alternatively, in order to balance the initial correlations with an initially non-vanishing interaction strength, we will consider here initial states with suitable correlations to the environment.
Such states will be obtained via an auxiliary construction which involves evolving an initially uncorrelated state for a sufficiently long time (a similar procedure was used in Ref.~\cite{PerezNadal08,PerezNadal08b} within the context of semiclassical gravity).
The system-enviroment correlations are then dynamically generated through the environmental interaction itself.
Our first examples of \emph{equilibrium preparation} will be the simplest mathematically,
while the final examples of \emph{non-equilibrium preparation} will be closer to actual laboratory experiments.

In all cases we will take the system and environment to be 
uncorrelated not at $t=0$ but in the infinite past.
\begin{equation}
\boldsymbol{\rho}_\mathrm{C}(-\infty) = \boldsymbol{\rho}_\mathrm{S}(-\infty) \otimes \boldsymbol{\rho}_\mathrm{E}(-\infty) \, , \label{eq:PreHistory}
\end{equation}
for some (possibly unimportant) system state $\boldsymbol{\rho}_\mathrm{S}(-\infty)$ and thermal $\boldsymbol{\rho}_\mathrm{E}(-\infty)$.
We then define the system Hamiltonian piecewise in time
\begin{equation}
\mathbf{H}_\mathrm{S}(t) = \left\{ \begin{array}{l@{\;\;\;\;}rcl} \mathbf{H}_+(t) & 0 & < & t \\ \mathbf{H}_- & t & < & 0 \end{array} \right. \, ,
\end{equation}
such that in past the system is allowed to equilibrate with the environment for an infinite time, which determines the correlated initial state at $t=0$.
The second-order master equation is then determined by
\begin{equation}
\left( \mathbf{A}_{nm}\! \diamond \mathbf{L}_m \right)\!(t) = \int_{\!-\infty}^t \!\!\! d\tau \, \alpha_{nm}(t,\tau) \, \left\{ \mathbf{G}_\mathrm{S}(t,\tau) \, \mathbf{L}_m(\tau) \right\} \, .
\end{equation}
To analyze the coefficients associated with the initially-correlated state, we will reduce them to a sum of coefficients for the auxiliary initially-uncorrelated state involving various time ranges.
First, we split the integration into two parts
\begin{equation}
\int_{\!-\infty}^t \!\!\! d\tau = \int_{0}^t \!\! d\tau + \int_{\!-\infty}^0 \!\!\! d\tau \, ,
\end{equation}
with the first integral depending only upon $\mathbf{G}_+(t,\tau)$ and corresponding to the uncorrelated coefficients.
Inserting  the product $\mathbf{G}_-(0,t) \, \mathbf{G}_-(t,0)$, which equals the identity, the second integral can be written as
\begin{equation}
\boldsymbol{\mathcal{M}}(t) \int_{\!-\infty}^0 \!\!\! d\tau \, \alpha_{nm}(t,\tau) \, \left\{ \mathbf{G}_-(t,\tau) \, \mathbf{L}_m(\tau) \right\} \, , \label{eq:Int2}
\end{equation}
given the operator
\begin{equation}
\boldsymbol{\mathcal{M}}(t)  \equiv  \mathbf{G}_+(t,0) \, \mathbf{G}_-(0,t) \, .
\end{equation}
The integral in Eq.~\eqref{eq:Int2} is then broken up into two parts
\begin{equation}
\int_{\!-\infty}^0 \!\!\! d\tau = \int_{\!-\infty}^t  \!\!\! d\tau - \int_{0}^t \!\! d\tau \, ,
\end{equation}
corresponding to the asymptotic and finite-time coefficients for an initially uncorrelated system driven by the time-independent preparation Hamiltonian $\mathbf{H}_-$.
Finally, our correlated coefficients can be expressed in terms of the uncorrelated coefficients as
\begin{align}
& \underbrace{\left( \mathbf{A}_{nm}\! \diamond \mathbf{L}_m \right)\!(t)}_\mathrm{correlated} = \underbrace{\left( \mathbf{A}_{nm}\! \diamond \mathbf{L}_m \right)_+\!(t)}_\mathrm{uncorrelated} \label{eq:Aprep} \\
& - \boldsymbol{\mathcal{M}}(t) \Bigl\{ \underbrace{\left( \mathbf{A}_{nm}\! \diamond \mathbf{L}_m \right)_-\!(t)}_\mathrm{jolt \; suppression}
 - \underbrace{\left( \mathbf{A}_{nm}\! \diamond \mathbf{L}_m \right)_-\!(\infty)}_\mathrm{preparation \; eraser} \Bigr\} \, , \nonumber
\end{align}
where the subscripted $\left( \mathbf{A} \diamond \mathbf{L} \right)_\pm$ coefficients are defined as
\begin{equation}
(\mathbf{A}_{nm}\! \diamond \mathbf{L}_m)_\pm(t) \equiv \int_0^t \!\! d\tau \, \alpha_{nm}(t,\tau) \, \left\{ \mathbf{G}_\pm(t,\tau) \, \mathbf{L}_m(\tau) \right\} \, .
\end{equation}
If the system frequencies are always small as compared to the cutoff,
we can inspect the early-time behavior (and jolts) by letting $\mathbf{G}_\pm(t) \approx \mathbf{1}$.
Then one can see that the first two terms of Eq.~\eqref{eq:Aprep} will precisely cancel in the early-time regime.
Therefore, the correlated initial states are jolt-free given sufficiently small system frequencies as compared to the cutoff: $\Omega \ll \Lambda$.
The final term in turn is such that in the late-time limit it precisely cancels the second term and erases all memory of $\mathbf{H}_-$.
Finally note that, quite trivially, if we choose $\mathbf{H}_+(t) = \mathbf{H}_-$,
then the first two terms cancel and we recover the equilibrium coefficients at any finite time.

\subsection{Equilibrium preparation}
To prepare an initial state in this approach, we choose the past Hamiltonian $\mathbf{H}_-$ such that its dynamics along with the environment interaction relaxes our system to the desired initial state:
\begin{eqnarray}
\lim_{t \to \infty} e^{t \, \boldsymbol{\mathcal{L}}_-(\infty)} \, \boldsymbol{\rho}_0 &=& \boldsymbol{\rho}_0 \, , \\
\boldsymbol{\mathcal{L}}_-(\infty) \, \boldsymbol{\rho}_0 &=& \mathbf{0} \, ,
\end{eqnarray}
where $\boldsymbol{\mathcal{L}}_-(\infty)$ is the stationary limit of the Liouvillian for a system with the past Hamiltonian as well as the coupling to the environment.

Our target state $\boldsymbol{\rho}_0$ will only be specified to zeroth order in the system-environment interaction. This is because for sufficiently long times (and in particular for the asymptotic equilibrium state) the diagonal elements of the reduced density matrix in the energy basis cannot be determined beyond zeroth order anyway when using the second-order perturbative master equation \cite{Accuracy}.
Due to unavoidable degeneracy present in all open-system dynamics, one actually requires components of the fourth-order master equation to calculate the full second-order solutions. The second-order master equation provides for all second-order dynamical quantities, such as frequency shifts, dissipation, diffusion and decoherence rates. We are concerned here with the induced jolts, which are dynamical quantities, and so this subtle point does not raise any additional problems for us.

\subsubsection{Preparation by decoherence}
For $\mathbf{L}_n$ all commuting with each other, one can force a general environment into $\ell$-state preparation via decoherence.
If the past Hamiltonian is deactivated, or more generally taken to commute with $\mathbf{L}_n$,
then since all system operators commute with each other, the master equation and its solutions will trivially result in a system which decoheres in the $\ell$-basis associated with the $\mathbf{L}_n$.
Thus, coefficients prepared in this manner are consistent with any initial state which is a completely incoherent mixture of $\ell$-states.
[Note that if $\boldsymbol{\rho}_\mathrm{S}(-\infty)$ corresponds to a pure eigenstate of the set $\{ \mathbf{L}_n \}$, this procedure simply adjusts the state of the environment, while system and environment remain unentangled.]

\subsubsection{Preparation by equilibration}
A finite-temperature environment allows mixed state preparation by equilibration.
Essentially one chooses the past Hamiltonian so that its thermal state (or some other steady state) is the desired initial state.
For a positive-temperature environment, at zeroth order one can prepare a (sufficiently) mixed state $\boldsymbol{\rho}_0$ with the past Hamiltonian $\mathbf{H}_- = - T \, \log(\boldsymbol{\rho}_0)$.
However, one must be careful that past system frequencies are small as compared to the high frequency jolts, otherwise this preparation will fail to remedy jolting.
One can work out that the \emph{adiabatic preparation} regime is given by
\begin{eqnarray}
\frac{p_\mathrm{max}}{p_\mathrm{min}} & \ll & e^{\beta \, \Lambda} \, ,
\end{eqnarray}
where $\Lambda$ is the jolt frequency and $p$ are the initial state probabilities
of preparation energy levels connected by $\mathbf{L}_n$.  (Clearly, for this method to work there can only be a finite number of such energy levels.)

\subsubsection{Preparation by freezing}
To prepare an initially pure state via equilibration at the order that we are working, one requires a zero-temperature environment for preparation by freezing.
Then one can choose any $\mathbf{H}_-$ with ground state $\boldsymbol{\rho}_0$.
It is important to emphasize that the reduced density matrix of the system corresponding to the ground state of the combined system will not be a pure state in general due to the entanglement between the system and the environment:
the free ground state of the system is a pure state, but the reduced density matrix of the open system is in general a mixed state beyond zeroth order in the system-environment coupling.
However, this point becomes irrelevant at the order that we are working since, as explained above, when using the second-order perturbative master equation to prepare the initial state by equilibration, one cannot meaningfully specify $\boldsymbol{\rho}_0$ beyond zeroth order.

\begin{figure}[h]
\centering
\includegraphics[width=7cm]{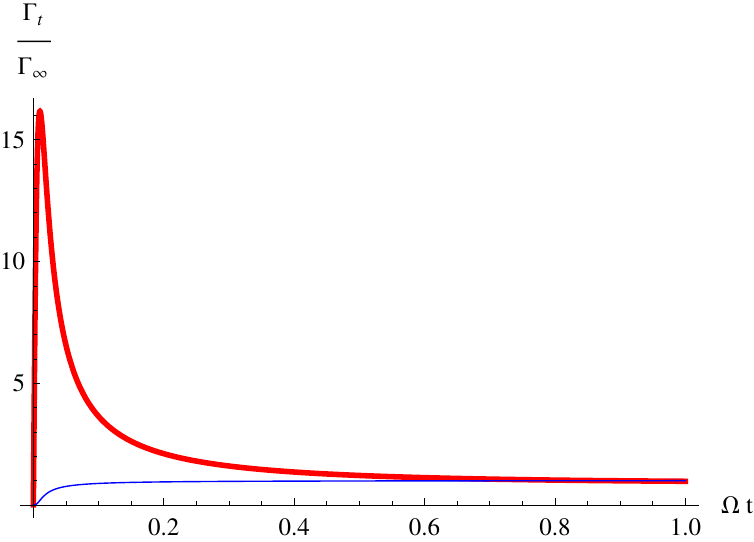}
\caption{Zero-temperature, ohmic decay rate for the \textcolor{red}{$\bullet$ unprepared} and \textcolor{blue}{$\cdot$ prepared} initial states of a two-level system with exponential cutoff frequency $\Lambda = 100 \, \Omega$.
In this case preparation by freezing was used to create an initially excited state.}
\label{fig:prepared}
\end{figure}


\subsection{Non-equilibrium preparation}
In order to consider situations closer to actual laboratory experiments, here we will first allow the system to equilibrate with the environment (as described in the previous subsection) and then choose some preparation Hamiltonian $\mathbf{H}_\mathrm{P}(t)$,
which would (in the absence of coupling to the environment) generate the desired initial state in some finite time $\tau_\mathrm{P}$.
One simply applies the master-equation coefficients in Eq.~\eqref{eq:Aprep} with future Hamiltonian
\begin{equation}
\mathbf{H}_+(t) = \left\{ \begin{array}{l@{\;\;\;\;}rcl} \mathbf{H}_0(t) & \tau_\mathrm{P} & < & t \\ \mathbf{H}_\mathrm{P}(t) & t & < & \tau_\mathrm{P} \end{array} \right. \, ,
\end{equation}
where $\mathbf{H}_0(t)$ is the desired post-preparation Hamiltonian.
All jolts will be avoided if $1/\tau_\mathrm{P} \ll \Lambda$: the introduction of a non-vanishing preparation frequency serves to tame the jolts and eliminate their high-cutoff sensitivity.

\subsubsection{State flipping}
A possible preparation Hamiltonian, which could model as a particular case Rabbi oscillations induced by an appropriate laser field acting on a two-level system, is the following:
\begin{equation}
\mathbf{H}_\mathrm{P} = \frac{\pi}{2 \tau_\mathrm{P}} \left( \ket{\psi_0}\!\!\bra{0} + \ket{0}\!\!\bra{\psi_0} \right) \, .
\end{equation}
Assuming that one has a zero-temperature environment and that the system is already equilibrated, driving the system with this Hamiltonian for a time $\tau_\mathrm{P}$ provides a relatively easy way of preparing an initial pure state $\ket{\psi_0}$. As discussed above, the reduced density matrix of the system will actually be a mixed state in general, because of the system-environment entanglement of the equilibrium state as well as the interaction to the environment while evolving the combined system during this additional finite preparation time. In fact, the preparation time $\tau_\mathrm{P}$ cannot be too long if we want the state of the system to be more or less close to $\ket{\psi_0}$.

\subsubsection{State swapping}

Let us consider a system which is initially equilibrated, without making any assumption as to the temperature of the environment.
We couple the system to an ancillary and analog system (equivalent Hilbert spaces) that is already prepared in the desired initial state.
The system of interest and ancilla are temporarily coupled in such a way that they swap states, for instance by means of the following block-matrix preparation Hamiltonian:
\begin{equation}
\mathbf{H}_\mathrm{P} = \frac{\pi}{2 \tau_\mathrm{P}} \left[ \begin{array}{cc} \mathbf{0} & \mathbf{1} \\ \mathbf{1} & \mathbf{0} \end{array} \right] \, .
\end{equation}
In the absence of coupling to the environment this would exactly swap the system and ancilla states in a time $\tau_\mathrm{P}$.
The same remarks as for state flipping concerning the purity and accuracy of the prepared state when taking into account the coupling to the environment also apply in this case.

\subsubsection{Other possibilities}
Within the second-order perturbative approach, generation of equilibrium correlations in a laboratory setting can always be calculated using Eq.~\eqref{eq:Aprep}.
One only needs to make sure that any additional state preparation does not rely upon large system energies as compared to the bath cutoff.
For instance, one can consider the preparation of Ref.~\cite{Anglin97}, which relies on ancillary degrees of freedom to drive the equilibrium state into a coherent superposition.
In fact, one could simply apply their own time-dependent Hamiltonian to our formulas as $\mathbf{H}_+(t)$ and obtain results consistent with theirs.


\acknowledgments

This work is supported partially by
NSF Grants PHY-0426696, PHY-0801368, DARPA grant
DARPAHR0011-09-1-0008 and the Laboratory of Physical Sciences.

\bibliography{bib}{}

\begin{thebibliography}{20}%
\makeatletter
\providecommand \@ifxundefined [1]{%
 \@ifx{#1\undefined}
}%
\providecommand \@ifnum [1]{%
 \ifnum #1\expandafter \@firstoftwo
 \else \expandafter \@secondoftwo
 \fi
}%
\providecommand \@ifx [1]{%
 \ifx #1\expandafter \@firstoftwo
 \else \expandafter \@secondoftwo
 \fi
}%
\providecommand \natexlab [1]{#1}%
\providecommand \enquote  [1]{``#1''}%
\providecommand \bibnamefont  [1]{#1}%
\providecommand \bibfnamefont [1]{#1}%
\providecommand \citenamefont [1]{#1}%
\providecommand \href@noop [0]{\@secondoftwo}%
\providecommand \href [0]{\begingroup \@sanitize@url \@href}%
\providecommand \@href[1]{\@@startlink{#1}\@@href}%
\providecommand \@@href[1]{\endgroup#1\@@endlink}%
\providecommand \@sanitize@url [0]{\catcode `\\12\catcode `\$12\catcode
  `\&12\catcode `\#12\catcode `\^12\catcode `\_12\catcode `\%12\relax}%
\providecommand \@@startlink[1]{}%
\providecommand \@@endlink[0]{}%
\providecommand \url  [0]{\begingroup\@sanitize@url \@url }%
\providecommand \@url [1]{\endgroup\@href {#1}{\urlprefix }}%
\providecommand \urlprefix  [0]{URL }%
\providecommand \Eprint [0]{\href }%
\providecommand \doibase [0]{http://dx.doi.org/}%
\providecommand \selectlanguage [0]{\@gobble}%
\providecommand \bibinfo  [0]{\@secondoftwo}%
\providecommand \bibfield  [0]{\@secondoftwo}%
\providecommand \translation [1]{[#1]}%
\providecommand \BibitemOpen [0]{}%
\providecommand \bibitemStop [0]{}%
\providecommand \bibitemNoStop [0]{.\EOS\space}%
\providecommand \EOS [0]{\spacefactor3000\relax}%
\providecommand \BibitemShut  [1]{\csname bibitem#1\endcsname}%
\let\auto@bib@innerbib\@empty
\bibitem [{\citenamefont {Hu}\ \emph {et~al.}(1992)\citenamefont {Hu},
  \citenamefont {Paz},\ and\ \citenamefont {Zhang}}]{HPZ92}%
  \BibitemOpen
  \bibfield  {author} {\bibinfo {author} {\bibfnamefont {B.~L.}\ \bibnamefont
  {Hu}}, \bibinfo {author} {\bibfnamefont {J.~P.}\ \bibnamefont {Paz}}, \ and\
  \bibinfo {author} {\bibfnamefont {Y.}~\bibnamefont {Zhang}},\ }\href@noop {}
  {\bibfield  {journal} {\bibinfo  {journal} {Phys. Rev. D}\ }\textbf {\bibinfo
  {volume} {45}},\ \bibinfo {pages} {2843} (\bibinfo {year}
  {1992})}\BibitemShut {NoStop}%
\bibitem [{\citenamefont {Hu}\ \emph {et~al.}(2004)\citenamefont {Hu},
  \citenamefont {Roura},\ and\ \citenamefont {Verdaguer}}]{HRV04}%
  \BibitemOpen
  \bibfield  {author} {\bibinfo {author} {\bibfnamefont {B.~L.}\ \bibnamefont
  {Hu}}, \bibinfo {author} {\bibfnamefont {A.}~\bibnamefont {Roura}}, \ and\
  \bibinfo {author} {\bibfnamefont {E.}~\bibnamefont {Verdaguer}},\ }\href@noop
  {} {\bibfield  {journal} {\bibinfo  {journal} {Phys. Rev. D}\ }\textbf
  {\bibinfo {volume} {70}},\ \bibinfo {pages} {044002} (\bibinfo {year}
  {2004})}\BibitemShut {NoStop}%
\bibitem [{\citenamefont {Grabert}\ \emph {et~al.}(1988)\citenamefont
  {Grabert}, \citenamefont {Schramm},\ and\ \citenamefont
  {Ingold}}]{Grabert88}%
  \BibitemOpen
  \bibfield  {author} {\bibinfo {author} {\bibfnamefont {H.}~\bibnamefont
  {Grabert}}, \bibinfo {author} {\bibfnamefont {P.}~\bibnamefont {Schramm}}, \
  and\ \bibinfo {author} {\bibfnamefont {G.~L.}\ \bibnamefont {Ingold}},\
  }\href@noop {} {\bibfield  {journal} {\bibinfo  {journal} {Phys. Rep.}\
  }\textbf {\bibinfo {volume} {168}},\ \bibinfo {pages} {115} (\bibinfo {year}
  {1988})}\BibitemShut {NoStop}%
\bibitem [{\citenamefont {Romero}\ and\ \citenamefont {Paz}(1997)}]{Romero97}%
  \BibitemOpen
  \bibfield  {author} {\bibinfo {author} {\bibfnamefont {L.~D.}\ \bibnamefont
  {Romero}}\ and\ \bibinfo {author} {\bibfnamefont {J.~P.}\ \bibnamefont
  {Paz}},\ }\href@noop {} {\bibfield  {journal} {\bibinfo  {journal} {Phys.
  Rev. A}\ }\textbf {\bibinfo {volume} {55}},\ \bibinfo {pages} {4070}
  (\bibinfo {year} {1997})}\BibitemShut {NoStop}%
\bibitem [{\citenamefont {Anglin}\ \emph {et~al.}(1997)\citenamefont {Anglin},
  \citenamefont {Paz},\ and\ \citenamefont {Zurek}}]{Anglin97}%
  \BibitemOpen
  \bibfield  {author} {\bibinfo {author} {\bibfnamefont {J.~R.}\ \bibnamefont
  {Anglin}}, \bibinfo {author} {\bibfnamefont {J.~P.}\ \bibnamefont {Paz}}, \
  and\ \bibinfo {author} {\bibfnamefont {W.~H.}\ \bibnamefont {Zurek}},\
  }\href@noop {} {\bibfield  {journal} {\bibinfo  {journal} {Phys. Rev. A}\
  }\textbf {\bibinfo {volume} {55}},\ \bibinfo {pages} {4041} (\bibinfo {year}
  {1997})}\BibitemShut {NoStop}%
\bibitem [{\citenamefont {Breuer}\ \emph {et~al.}(2001)\citenamefont {Breuer},
  \citenamefont {Kappler},\ and\ \citenamefont {Petruccione}}]{Breuer01b}%
  \BibitemOpen
  \bibfield  {author} {\bibinfo {author} {\bibfnamefont {H.~P.}\ \bibnamefont
  {Breuer}}, \bibinfo {author} {\bibfnamefont {B.}~\bibnamefont {Kappler}}, \
  and\ \bibinfo {author} {\bibfnamefont {F.}~\bibnamefont {Petruccione}},\
  }\href@noop {} {\bibfield  {journal} {\bibinfo  {journal} {Ann. Phys.}\
  }\textbf {\bibinfo {volume} {291}},\ \bibinfo {pages} {36} (\bibinfo {year}
  {2001})}\BibitemShut {NoStop}%
\bibitem [{\citenamefont {van Kampen}\ and\ \citenamefont
  {Oppenheim}(1997)}]{Kampen97}%
  \BibitemOpen
  \bibfield  {author} {\bibinfo {author} {\bibfnamefont {N.~G.}\ \bibnamefont
  {van Kampen}}\ and\ \bibinfo {author} {\bibfnamefont {I.}~\bibnamefont
  {Oppenheim}},\ }\href@noop {} {\bibfield  {journal} {\bibinfo  {journal} {J.
  Stat. Phys.}\ }\textbf {\bibinfo {volume} {87}},\ \bibinfo {pages} {1325}
  (\bibinfo {year} {1997})}\BibitemShut {NoStop}%
\bibitem [{\citenamefont {Breuer}\ and\ \citenamefont
  {Petruccione}(2001)}]{Breuer01}%
  \BibitemOpen
  \bibfield  {author} {\bibinfo {author} {\bibfnamefont {H.~P.}\ \bibnamefont
  {Breuer}}\ and\ \bibinfo {author} {\bibfnamefont {F.}~\bibnamefont
  {Petruccione}},\ }\href@noop {} {\bibfield  {journal} {\bibinfo  {journal}
  {Phys. Rev. A}\ }\textbf {\bibinfo {volume} {63}},\ \bibinfo {pages} {032102}
  (\bibinfo {year} {2001})}\BibitemShut {NoStop}%
\bibitem [{\citenamefont {Strunz}\ and\ \citenamefont {Yu}(2004)}]{Strunz04}%
  \BibitemOpen
  \bibfield  {author} {\bibinfo {author} {\bibfnamefont {W.~T.}\ \bibnamefont
  {Strunz}}\ and\ \bibinfo {author} {\bibfnamefont {T.}~\bibnamefont {Yu}},\
  }\href@noop {} {\bibfield  {journal} {\bibinfo  {journal} {Phys. Rev. A}\
  }\textbf {\bibinfo {volume} {69}},\ \bibinfo {pages} {052115} (\bibinfo
  {year} {2004})}\BibitemShut {NoStop}%
\bibitem [{\citenamefont {Pollard}\ \emph {et~al.}(1997)\citenamefont
  {Pollard}, \citenamefont {Felts},\ and\ \citenamefont
  {Friesner}}]{Pollard97}%
  \BibitemOpen
  \bibfield  {author} {\bibinfo {author} {\bibfnamefont {W.~T.}\ \bibnamefont
  {Pollard}}, \bibinfo {author} {\bibfnamefont {A.~K.}\ \bibnamefont {Felts}},
  \ and\ \bibinfo {author} {\bibfnamefont {R.~A.}\ \bibnamefont {Friesner}},\
  }\href@noop {} {\bibfield  {journal} {\bibinfo  {journal} {Adv. Chem. Phys.}\
  }\textbf {\bibinfo {volume} {93}},\ \bibinfo {pages} {77} (\bibinfo {year}
  {1997})}\BibitemShut {NoStop}%
\bibitem [{\citenamefont {Carmichael}(1999)}]{Carmichael99}%
  \BibitemOpen
  \bibfield  {author} {\bibinfo {author} {\bibfnamefont {H.~J.}\ \bibnamefont
  {Carmichael}},\ }\href@noop {} {\emph {\bibinfo {title} {Statistical Methods
  in Quantum Optics I}}}\ (\bibinfo  {publisher} {Springer},\ \bibinfo
  {address} {New York},\ \bibinfo {year} {1999})\BibitemShut {NoStop}%
\bibitem [{\citenamefont {Breuer}\ and\ \citenamefont
  {Petruccione}(2002)}]{Breuer02}%
  \BibitemOpen
  \bibfield  {author} {\bibinfo {author} {\bibfnamefont {H.~P.}\ \bibnamefont
  {Breuer}}\ and\ \bibinfo {author} {\bibfnamefont {F.}~\bibnamefont
  {Petruccione}},\ }\href@noop {} {\emph {\bibinfo {title} {The Theory of Open
  Quantum Systems}}}\ (\bibinfo  {publisher} {Oxford University Press},\
  \bibinfo {address} {Oxford},\ \bibinfo {year} {2002})\BibitemShut {NoStop}%
\bibitem [{\citenamefont {Kampen}(2007)}]{Kampen07}%
  \BibitemOpen
  \bibfield  {author} {\bibinfo {author} {\bibfnamefont {N.~G.~V.}\
  \bibnamefont {Kampen}},\ }\href@noop {} {\ \textbf {\bibinfo {volume} {34}},\
  \bibinfo {pages} {245} (\bibinfo {year} {2007})}\BibitemShut {NoStop}%
\bibitem [{\citenamefont {Fleming}\ and\ \citenamefont {Hu}(2011)}]{QOS}%
  \BibitemOpen
  \bibfield  {author} {\bibinfo {author} {\bibfnamefont {C.~H.}\ \bibnamefont
  {Fleming}}\ and\ \bibinfo {author} {\bibfnamefont {B.~L.}\ \bibnamefont
  {Hu}},\ }\href@noop {} {\enquote {\bibinfo {title} {The evolution of general
  systems in non-markovian environments},}\ } (\bibinfo {year} {2011}),\
  \bibinfo {note} {\emph{in preparation}}\BibitemShut {NoStop}%
\bibitem [{\citenamefont {Kubo}(1957)}]{Kubo57}%
  \BibitemOpen
  \bibfield  {author} {\bibinfo {author} {\bibfnamefont {R.}~\bibnamefont
  {Kubo}},\ }\href@noop {} {\bibfield  {journal} {\bibinfo  {journal} {J. Phys.
  Soc. Japan}\ }\textbf {\bibinfo {volume} {12}},\ \bibinfo {pages} {570}
  (\bibinfo {year} {1957})}\BibitemShut {NoStop}%
\bibitem [{\citenamefont {Martin}\ and\ \citenamefont
  {Schwinger}(1959)}]{Martin59}%
  \BibitemOpen
  \bibfield  {author} {\bibinfo {author} {\bibfnamefont {P.~C.}\ \bibnamefont
  {Martin}}\ and\ \bibinfo {author} {\bibfnamefont {J.}~\bibnamefont
  {Schwinger}},\ }\href@noop {} {\bibfield  {journal} {\bibinfo  {journal}
  {Phys. Rev.}\ }\textbf {\bibinfo {volume} {115}},\ \bibinfo {pages} {1342}
  (\bibinfo {year} {1959})}\BibitemShut {NoStop}%
\bibitem [{\citenamefont {Fleming}\ \emph {et~al.}(2010)\citenamefont
  {Fleming}, \citenamefont {Roura},\ and\ \citenamefont {Hu}}]{QBM}%
  \BibitemOpen
  \bibfield  {author} {\bibinfo {author} {\bibfnamefont {C.~H.}\ \bibnamefont
  {Fleming}}, \bibinfo {author} {\bibfnamefont {A.}~\bibnamefont {Roura}}, \
  and\ \bibinfo {author} {\bibfnamefont {B.~L.}\ \bibnamefont {Hu}},\
  }\href@noop {} {\enquote {\bibinfo {title} {Exact analytical solutions to the
  master equation of quantum brownian motion for a general environment},}\ }
  (\bibinfo {year} {2010}),\ \bibinfo {note} {\emph{accepted for publication,
  Ann. Phys. (NY)}},\ \Eprint {http://arxiv.org/abs/1004.1603} {arXiv:1004.1603
  [quant-ph]} \BibitemShut {NoStop}%
\bibitem [{\citenamefont {P\'{e}rez-Nadal}\ \emph
  {et~al.}(2008{\natexlab{a}})\citenamefont {P\'{e}rez-Nadal}, \citenamefont
  {Roura},\ and\ \citenamefont {Verdaguer}}]{PerezNadal08}%
  \BibitemOpen
  \bibfield  {author} {\bibinfo {author} {\bibfnamefont {G.}~\bibnamefont
  {P\'{e}rez-Nadal}}, \bibinfo {author} {\bibfnamefont {A.}~\bibnamefont
  {Roura}}, \ and\ \bibinfo {author} {\bibfnamefont {E.}~\bibnamefont
  {Verdaguer}},\ }\href@noop {} {\bibfield  {journal} {\bibinfo  {journal}
  {Phys. Rev. D}\ }\textbf {\bibinfo {volume} {77}},\ \bibinfo {pages} {124033}
  (\bibinfo {year} {2008}{\natexlab{a}})}\BibitemShut {NoStop}%
\bibitem [{\citenamefont {P\'{e}rez-Nadal}\ \emph
  {et~al.}(2008{\natexlab{b}})\citenamefont {P\'{e}rez-Nadal}, \citenamefont
  {Roura},\ and\ \citenamefont {Verdaguer}}]{PerezNadal08b}%
  \BibitemOpen
  \bibfield  {author} {\bibinfo {author} {\bibfnamefont {G.}~\bibnamefont
  {P\'{e}rez-Nadal}}, \bibinfo {author} {\bibfnamefont {A.}~\bibnamefont
  {Roura}}, \ and\ \bibinfo {author} {\bibfnamefont {E.}~\bibnamefont
  {Verdaguer}},\ }\href@noop {} {\bibfield  {journal} {\bibinfo  {journal}
  {Class. Quant. Grav.}\ }\textbf {\bibinfo {volume} {25}},\ \bibinfo {pages}
  {154013} (\bibinfo {year} {2008}{\natexlab{b}})}\BibitemShut {NoStop}%
\bibitem [{\citenamefont {Fleming}\ and\ \citenamefont
  {Cummings}(2010)}]{Accuracy}%
  \BibitemOpen
  \bibfield  {author} {\bibinfo {author} {\bibfnamefont {C.~H.}\ \bibnamefont
  {Fleming}}\ and\ \bibinfo {author} {\bibfnamefont {N.~I.}\ \bibnamefont
  {Cummings}},\ }\href@noop {} {\enquote {\bibinfo {title} {On the accuracy of
  perturbative master equations},}\ } (\bibinfo {year} {2010}),\ \bibinfo
  {note} {\emph{submitted to Phys. Rev. E}},\ \Eprint
  {http://arxiv.org/abs/1010.5025} {arXiv:1010.5025 [quant-ph]} \BibitemShut
  {NoStop}%
\end{thebibliography}%
\bibliographystyle{apsrev4-1}

\end{document}